# Regenerative Soot-IX: $C_3$ as the dominant, stable carbon cluster in high pressure sooting discharges

Sohail Ahmad Janjua[1], M. Ahmad[1], S. D. Khan[1], R. Khalid[1], A. Aleem[1] and Shoaib Ahmad[1,2]

[1]*PINSTECH, P.O. Nilore, Islamabad, Pakistan*
[2]*National Centre for Physics, Quaid-i-Azam University Campus, Shahdara Valley, Islamabad, 44000, Pakistan*

Email: sahmad.ncp@gmail.com

**Abstract**
Results are presented that have been obtained while operating the graphite hollow cathode duoplasmatron ion source in dual mode under constant discharge current. This dual mode operation enabled us to obtain the mass and emission spectra simultaneously. In mass spectra $C_3$ is the main feature but $C_4$ and $C_5$ are also prominent, whereas in emission spectra $C_2$ is also there and its presence shows that it is in an excited state rather than in an ionic state. These facts provide evidence that $C_3$ is produced due to the regeneration of a soot forming sequence and leave it in ionic state. $C_3$ is a stable molecule and the only dominant species among the carbon clusters that survives in a regenerative sooting environment at high-pressure discharges.

**Introduction**

The aim of the present investigation is to study the effect of pressure of the discharge support gas on the formation and fragmentation mechanisms of the carbon clusters produced in regenerative sooting discharges. The graphite hollow cathode (HC) sources are very efficient in creating sooting environments whose regeneration can lead to the formation of carbon clusters [1]. Many techniques have been reported for the production of C clusters [2–6]. The kind of source being reported in this communication can be operated in dual modes, i.e. HC discharge mode and duoplasmatron mode with Ar as support gas to initiate the discharge. A similar source with only HC discharge and Ne as support gas has been studied earlier [7]. However, the basic difference between our mode of operation and the earlier lies in the higher levels of ionization with Ne discharges in graphite HCs surrounded by cusp magnetic fields and the present one where Ar discharge occurs without a cusp magnetic field around the HC. Also, up to an order of magnitude higher Ar pressure is used in our present study. The sooting discharge is initially created and sustained in the graphite hollow cathode whose inner surface is sputtered by the energetic Ar ions. The sputtered species contain $C_1$, $C_2$, $C_3$ and higher C clusters which start to actively participate in all of the discharge activities including inter and intra-species collisions. The initially pure Ar discharge gets transformed into a carbonaceous discharge with variable densities of different C clusters [8]. The transformation of the Ar discharge to the one with $Ar^+$ and C clusters is accompanied by the substitution of the pure graphite cathode surface with one that is coated with multiple layers of carbon clusters. This agglomerate of C clusters subsequently becomes the new cathode surface and it is this surface whose regeneration as a function of the glow discharge pressure is the subject of the present study.

The initial and most significant stage in the production of regenerative soot depends on the initiation of the glow discharge of the noble gas as a function of source parameters like the geometry, discharge voltage and current. The sputtering introduces carbon into the plasma. Formation of carbon clusters $C_x$ ($x \geq 2$) takes place in plasmas as well as on the cathode walls, where all neutral and excited species are deposited. The C clusters produce a sooted graphite surface, which takes over the underlying graphite as the effective HC. All subsequent emissions, i.e. electrons and other constituents from the cathode, are in fact from this sooted layer [9]. This process has similarities with the soot production by arc discharge of Kratschmer *et al* [10]; however, the main difference is in the continuous operation of the sooted source. In our case this recycles the soot from which C atoms, ions and clusters are emitted into the discharge. This process is called the regeneration of the soot by which one can manipulate the overall constitution of the soot by adjusting the discharge parameters. The regenerative sooting environment that leads to the formation of C-clusters has been studied with two diagnostics techniques of mass spectrometry and emission spectroscopy. A compact and permanent magnet based $E \times B$ velocity filter is used for carbon cluster diagnostics that has already been reported [11].

A multi-component glow discharge whose composition is continuously changing cannot be in local thermodynamic equilibrium (LTE). It may switch from non-LTE to partial LTE and vice versa. If the plasma is in non-LTE i.e. population inversion (PI), defined by $[N_m g_n / N_n g_m] > 1$, then we cannot define any temperature using the Boltzmann distribution. If there is a partial LTE i.e. PI < 1, then we can only calculate the excitation temperature $T_{exc}$ for those pairs of levels of a species, which falls on a common level, rather than a typical electron temperature $T_e$ for the plasma. The ratios of the upper level density ($N_m$) and lower level density ($N_n$) for respective transitions to the same level $r$ and the line intensities $I_{mr}$ and $I_{nr}$ are related by $I_{mr} = N_m A_{mr} h\nu_{mr}$ and similarly for $I_{nr}$, where $h\nu_{mr}$ is the energy difference between m and r levels and $A_{mr}$ is the Einstein transition probability for spontaneous emission. The two intensities are used for the evaluation of the relative level densities $N_m/N_n$ given by

$N_m/N_n = (g_m/g_n) \exp[-(E_m - E_n)/kT_{exc}]$ [12, 13],

where $g_m$ and $g_n$ are the statistical weights, $E_m$ and $E_n$ are energies of the respective levels. $T_{exc}$ is the excitation temperature of the atomic or the vibration temperature of the molecular species as the case may be.

An ion is passed un-deflected through the $E \times B$ velocity filter if it satisfies the condition that its velocity $v_o = E_o/B_o$. A velocity spectrum is taken by varying the compensating electric field $E_o$ which is related to mass as $M_o \propto (E_o)^{-2}$. Energy of ions after accelerating it to higher energies, in our case, is up to 2.5 kV. Increasing energy implies a larger spread between the ions of different masses; this effect is especially useful for heavier masses at the lower velocity end of the spectrum. The net effect is the improved resolution.

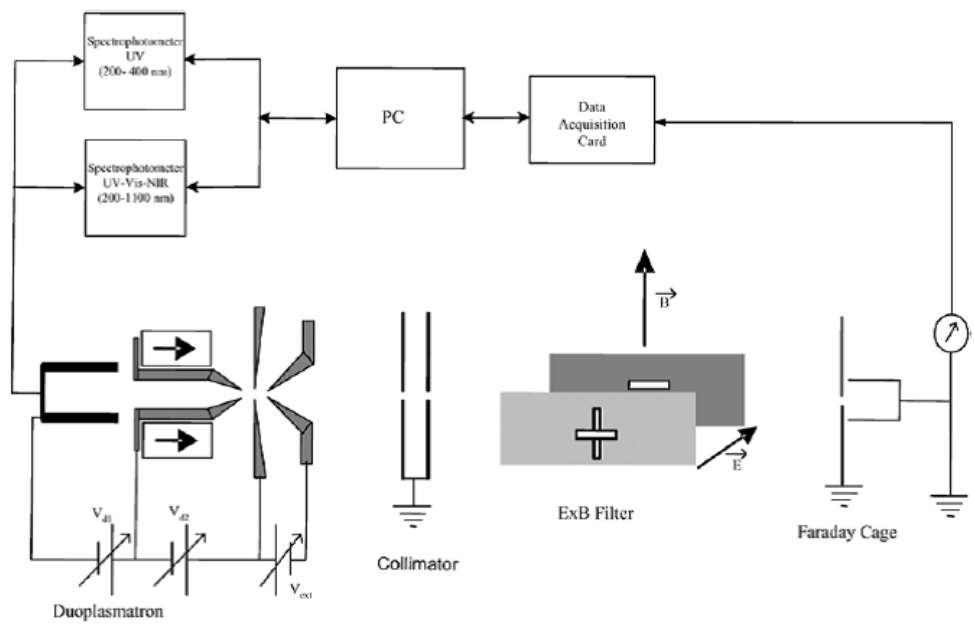

**Figure 1.** The schematic diagram of the experimental setup used for the study of effects of pressure of support gas on the relative intensities of the carbon clusters produced in carbonaceous plasmas. The first part of the setup is the duoplasmatron ion source operating in two independent modes; one with only the graphite HC discharge mode in which we vary the pressure of support gas Ar from 1 to 500 mbar and operate it for a few minutes and the other is the duoplasmatron mode operating at 0.6 mbar and constant discharge current $i_{dis}$ = 100 mA for emission and mass analysis after each successive pressure variation in the HC discharge mode. The magnetic field is also indicated with arrows. The numbered items are described in detail in the experiment.

## Experimental

The schematic diagram of the experimental setup is shown in figure 1. This setup is used to study the physical processes involved in graphite HC discharges as a function of Ar pressure from 1 to 500 mbar. The first part of the setup in figure 1 shows the geometry of the graphite HC duoplasmatron ion source used for the production of carbon clusters. In our design we use a graphite HC followed by an intermediate electrode (IE) and anode-A made of iron. The axial magnetic field is applied through IE, which is confined in between anode-A and the IE region of much smaller dimensions than the HC-IE region; hence producing intense ionization and initiating an arc discharge between the IE-A region, whereas there is a glow discharge in the HC-IE region. The ion source is operated in two independent modes, i.e. the HC discharge mode and the duoplasmatron mode.

## Hollow cathode discharge mode

First of all we operate the ion source in the HC discharge mode, i.e. the pure glow discharge mode in which we vary the pressure of support gas Ar from 1 to 500 mbar and operate it at any desired pressure for a few minutes. To switch the source to the HC discharge mode we short circuit the IE and A, converting both IE and A into the anode facing graphite HC. This glow discharge initiates sputtering of the graphite cathode by energetic Ar ions and introduces excited and ionic C monatomic, diatomic, tri-atomic and higher C clusters into the initially pure Ar discharge. Such a discharge has earlier been described as the regenerative sooting discharge and the variations of its C cluster composition as a function of discharge current and pressure have been studied for He and Ne in the low pressure range from 0.1 to 20 mbar [1].

## Duoplasmatron mode

After operating the source in the HC discharge mode for regenerating the soot at a given Ar pressure the source is switched to the duoplasmatron mode, which takes a minute and is operated at 0.6 mbar and constant discharge current $i_{dis}$ = 100 mA for emission and mass analysis after each successive pressure variation in the HC discharge mode. The emission spectroscopy and velocity filter analyzer are employed for the identification of the carbonaceous discharge environment in a duoplasmatron mode. The mass analysis of the extracted species is done by using a permanent magnet based $E \times B$ velocity filter, whereas the emission spectroscopy of the discharges is performed by using two Ocean Optics spectrophotometers UV (200–400 nm) and UV–Vis–NIR (200–1100 nm). Both the data are taken simultaneously. The UV–Vis–NIR spectrophotometer has a resolution of ∼1 nm with a broad wavelength range in which one can get a good picture of the overall discharge while the UV spectrophotometer has a higher resolution of ∼0.16 nm with a shorter range to especially identify and analyze the C2 swan band (430–570 nm).

We studied the excited levels of the discharge species in the atomic and ionic stages in the plasma as a function of the state of sooting of the graphite HC surface with these spectrophotometers. Mass analysis has been performed on positive ions extracted from the source by applying 2.5 kV. The extracted beam is passed through the $E \times B$ velocity filter, after collimating it to a beam of half angle of ∼2.3°. The multi-species ions, in the beam, are deflected by a fixed magnetic field of 0.35 T at angles depending upon their respective velocities at a constant energy of 2.5 keV. This deflection is compensated by a varying electric field for a straight through beam and detected by a faraday cage at a distance of 500 mm from the exit of the $E \times B$ velocity filter. The signal goes to the electrometer that measures the current in the picoampere (pA) range and finally to the data acquisition software which takes 3 min for each full spectrum in the range of the applied electric field. The process is repeated by varying the pressure to higher values after switching the source to the HC discharge mode and then to the duoplasmatron mode for taking mass and emission spectra and it is continued up to a pressure of 500 mbar.

It is pointed out here that although we operate the source in the duoplasmatron mode for emission as well as mass analysis at constant pressure, it possesses all the residual effects of the pressure changes, which have taken place in the HC discharge mode. Also the time taken to switch from one mode to another mode is about a minute or so, which ensures that the source does not cool down to disturb the pressure effects on the soot that has been regenerated in the HC discharge mode and its composition measured in the duoplasmatron mode.

# Results and discussion

The formative stages of carbon clusters in carbonaceous plasma are studied by using a specially designed duoplasmatron ion source. The mass analysis alone does not give a complete picture of the dynamics of events leading to the emission of clusters from a sooting discharge; therefore, we take emission and mass spectra simultaneously. To accomplish this task we operate the source in two modes—HC discharge mode and duoplasmatron mode. The dynamic mode operation of the source is for the first time reported in the present study.

Carbon in the atomic and molecular forms is introduced by the kinetic and potential sputtering of the walls of the graphite HC, which gets coated with the C clusters. The agglomerate of the C clusters on the walls creates a modified cathode surface and hence the subsequent discharges at increasing pressures provide an ideal environment for the formation and sustenance of stable clusters. Due to the regenerative processes in such a sooting environment there will be competition between the formation and fragmentation of clusters and the most stable ones under the given conditions are most likely to survive. Mass spectrometry supported by the emission spectroscopy of the sooting discharge resolves some of the issues relating to the diagnosis of the conditions of clusters formation with its constituents comprising of atomic and ionic Ar and various atomic and molecular C species.

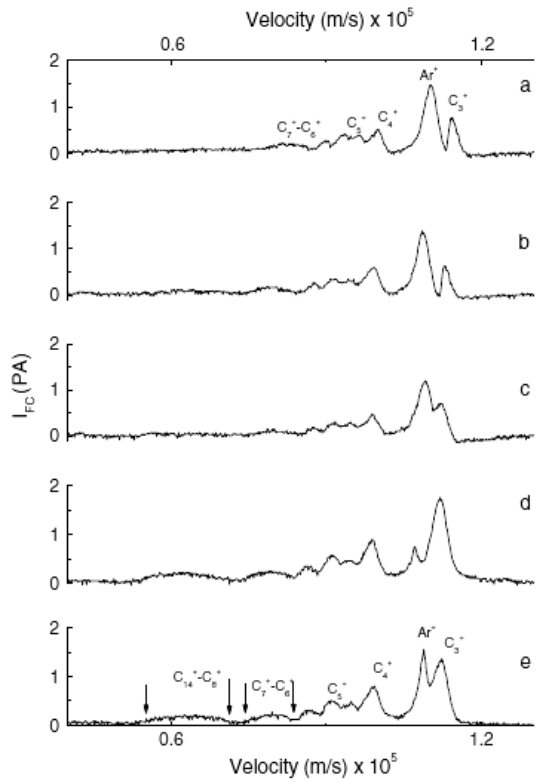

**Figure 2.** The velocity spectra of the positive charged clusters emitted from the source operated in the duoplasmatron mode at 0.6 mbar of Ar pressure, discharge current $i_{dis}$ = 100 mA, discharge voltage $V_{dis}$ = 0.75 kV and the extraction voltage $V_{ext}$ = 2.5 kV after each successive pressure variation from 1, 10, 100, 200 and 500 mbar in the HC discharge mode respectively.

## Mass spectrometry

The mass spectra of the positive charged clusters emitted from the source operated in the duoplasmatron mode at 0.6 mbar of Ar pressure, discharge current $i_{dis}$ = 100 mA, discharge voltage $V_{dis}$ = 0.75 kV and the extraction voltage $V_{ext}$ = 2.5 kV has certain unique characteristics. In general all spectra are dominated by $C_3$ and Ar along with smaller yields of $C_4$ to $C_{14}$. Figures 2(*a*)–(*e*) show the five spectra corresponding to pressures of 1, 10, 100, 200 and 500 mbar, respectively. The tendency of gradual increase in the yields of higher clusters with the increasing pressure is clearly evident. The second important observation is that $C^+$ is the major surviving species throughout the spectra.

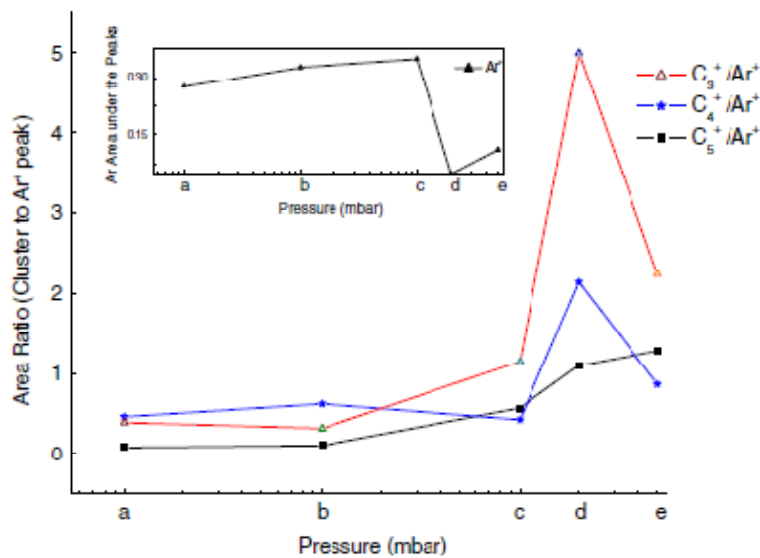

**Figure 3.** The area under the peaks of $C_3^+$, $C_4^+$ and $C_5^+$ normalized to the $Ar^+$ peak areas as a function of Ar pressure $P_{Ar}$ in the HC discharge mode. The graph shows that the yields of $C_3^+$ and $C_4^+$ are maximum around d (200 mbar) while $C^+_5$ has an increasing trend with the pressure. It is also clear from the spectrum that around d all the clusters have the maximum yield.

The area under the peaks of $C_3^+$, $C_4^+$ and $C_5^+$ normalized to the $Ar^+$ peak areas taken in the uno-plasmatron mode are plotted in figure 3 as a function of the soot forming sequence a, b, c, d, and e corresponding to Ar pressures of 1, 10, 100, 200 and 500 mbar, respectively, in the HC discharge mode. The insert in figure 3, which is the area under the Ar peaks shows the behavior of Ar throughout the operation of the ion source. It is clear from the graph that the higher the pressure the higher is the yield of the larger clusters. The yield of $C_3^+$ and $C_4^+$ is the maximum around (d) i.e. 200 mbar, shown in figure 2(*d*), while $C_5^+$ has an increasing trend with the pressure. This means that at higher pressure the sputtering with $Ar^+$ from the soot coated surface of the cathode increases, which ultimately increases the probability of the production of larger clusters. It is also clear from the spectrum that around 200 mbar all the clusters have the maximum yield, which indicates that it is the optimal value of pressure, and provide a conducive environment to synthesize the larger clusters and possibly even the fullerenes. Heath [14] has shown that nucleation in carbon vapor in connection with the fullerene formation indicates that intermediate complex $C_3^+$ is long-lived, which improves the formation probability of larger clusters.

## State of the atomic and ionized species

Figure 4(i) shows the spectrum of the ion source at gas pressure $P_{Ar}$ = 0.6 mbar, discharge current $i_{dis}$ = 100 mA and discharge voltage $V_{dis}$ = 0.75 kV. The spectrum shows all the respective Ar lines both atomic and ionic. Figure 4(ii) shows the plot of the typical excited Ar, i.e. Ar I level density calculated from the intensity of the line at $\lambda$ = 696 nm for the whole range of soot forming sequence. It shows a dip around (d) i.e. 200 mbar. The level density of ionized Ar (Ar II) at $\lambda$ = 476.4 nm is shown in figure 4(iii) as filled circles and the upper vibrational level density of $C_2$ at $\lambda$ = 473 nm as unfiled circles in Fig. 4(iv). Both level densities show a modest increase with pressure. These level densities show a competition mechanism between the collisional excitations between the energizing species of the discharge i.e., electrons and the high potential energy metastable atoms and ions. It leads to two mechanisms, i.e. electron collisional-induced processes and inter-particle collisions. The collisions between the excited, sputtered species may be responsible for the formation of more C3.

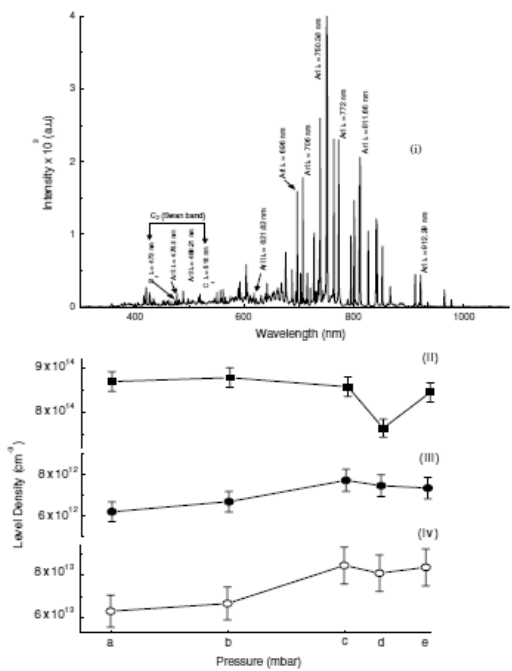

**Figure 4.** (i) The emission spectrum taken from the Ar discharge in the duoplasmatron mode is shown in the entire range of the spectrum. The spectrum shows familiar atomic lines and a host of other emission lines some of which are indicated. Many lines of Ar I and Ar II are observed as well as $C_2$'s Swan band in the 450–570 nm range in which the band heads are at 516 and 473 nm. The number densities of the excited levels for Ar I, Ar II, and $C_2$ are plotted as a function of the soot forming sequence in (ii), (iii) and (iv), respectively.

## $T_{exc}$ in the carbonaceous discharge

The emission data can be used to evaluate the discharge parameters like the excitation temperatures $T_{exc}$ for the atomic and ionic species such as AR I and AR II from level densities of the emission lines that have a common lower level. These $T_{exc}$ are plotted for Ar I in figure 5(iii) as a function of the soot forming sequence in the HC discharge mode. In figure 5(i) we have the Ar energy level diagram of the levels at $\lambda = 696$ nm and $\lambda = 772$ nm used for the calculation of the excitation temperature. There is a moderate decrease in $T_{exc}$ around (d), i.e. 200 mbar and an average value of $T_{exc}$ is $\sim$2350 K. Corresponding $T_{exc}$ for Ar II is $\sim$2150 K. In figure 5(ii) the $C_2$ energy level diagram of the levels at $\lambda = 516$ nm and $\lambda = 473$ nm is shown. The vibrational temperature $T_{vib}$ of $C_2$ is obtained from the lowest level transitions of the Swan band and is shown in figure 5(iv). The average value of $T_{vib}$ is $\sim$3750 K. It also shows a peak around (c), i.e. 100 mbar.

These three temperatures indicate a low temperature plasma ($\sim$2000–4000 K). These different temperatures show that our system is not in LTE, so different species have different temperatures. The origin of the data point error bars can be explained in terms of the statistical variation of the number density from where we calculate the errors. In figure 5(iii) the $T_{exc}$ decreases with pressure because as we increase the pressure in the HC discharge mode the more the cathode wall sputtering transforms the initial $Ar^+$-dominant discharge into a carbonaceous one due to soot formation till the soot forming sequence reaches (d) i.e. 200 mbar. Above this value the heavier clusters start to form and hence their total number decreases and the number of Ar atoms again start increasing. The same argument explains the peak of figure 5(iv).

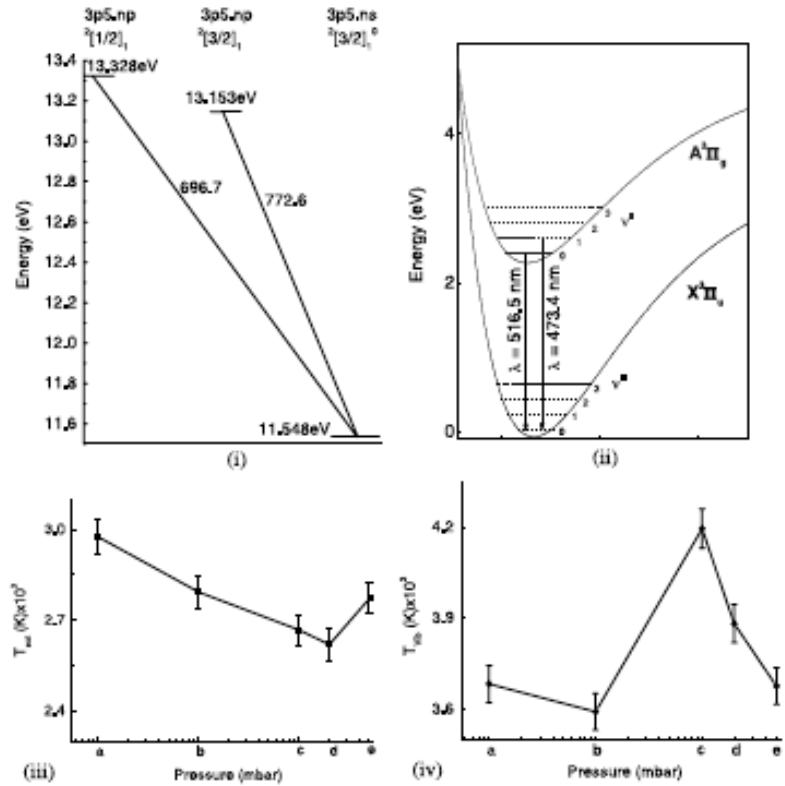

**Figure 5.** The energy level diagram of Ar I and $C_2$ show the representative transition from (696–772 nm) and (473–516 nm) in (i) and (ii), respectively. Also the calculated excitation temperature $T_{exc}$ of Ar I and the vibration temperature of $C_2$ of the corresponding levels are plotted against pressure $P_{Ar}$, as shown in (iii) and (iv), respectively.

## Conclusion

It has been shown that the simultaneous mass spectrometry of the clusters extracted from the source and the emission spectroscopy of the carbonaceous discharge yields important information about the dynamical process of the formation and fragmentation of clusters within the sooting discharge as a function of the number density of Ar. We conclude that $C^+$ is a stable molecule and the dominant species among the carbon clusters that survives in high- pressure discharges. It is also concluded that $C_3^+$ is mainly produced as the by-product of the regeneration sequence. This fact is further supported by the presence of excited $C_2$. Also the different temperatures of Ar and $C_2$ suggest that our discharge is in partial LTE. The density of linear clusters increases with Ar pressure; which is a signature of transformation from linear structures to ring structures that may also lead to cage closure. The size and type of carbon clusters formation and fragmentation has been seen in our present investigation to depend heavily on the collisional processes with the discharge gas, i.e. Ar.

## References

[1] Ahmad S 2002 *Eur. Phys. J.* D **18** 309

[2] Rohlfing E A, Cox D M and Kaldor A 1984 *J. Chem. Phys.* **81** 3322

[3] Kroto H W, Heath J R, O'Brien S C, Curl R F and Smalley R E 1985 *Nature* **318** 162

[4] Kratschmer W, Lamb L D, Fostiropoulous K and Huffman D R 1990 *Nature* **347** 354

[5] Ugarte D 1992 *Nature* **359** 707

[6] Iijima S, Ajayan P M and Ichihashi T 1992 *Phys. Rev. Lett.* **69** 3100

[7] Ahmad S, Ahmad B and Riffat T 2001 *Phys. Rev.* E **64** 0264 [8] Ahmad S 1999 *Phys. Lett.* A **261** 327

[9] Ahmad S, Qayyum A, Akhtar M N and Riffat T 2000 *Nucl. Instrum. Methods* B **171** 552

[10] Kratschmer W, Fostiropoulous K and Huffman D R 1990 *Chem. Phys. Lett.* **170** 167

[11] Ahmad S, Ahmad B, Qayyum A and Akhtar M N 2000 *Nucl. Instrum. Methods* A **452** 371

[12] Sobel'man I I 1972 *Introduction to the Theory of Atomic Spectra* (Oxford: Pergamon) chapter 9

[13] Vitense E B 1993 *Introduction to Stellar Atmospheres* vol 2 (Cambridge: Cambridge University Press) chapter 13

[14] Heath J R, Hammond G S and Kuck V J 1992 *Fullerenes* (*ACS Symposium Series* vol 481) (Washington, DC: American Chemical Society)